\documentclass[11pt]{article}

\begin{document}
\sloppy

\setcounter{page}{1}

\title{
Minimizing makespan in flowshop with time lags
}

\author{
Julien Fondrevelle (Speaker)
   \thanks{ {\tt julien.fondrevelle@loria.fr}.
   MACSI Project LORIA-INRIA Lorraine,
   Ecole des Mines de Nancy,
   Parc de Saurupt, 54042 Nancy, France.
   }
\and
Ammar Oulamara 
   \thanks{ {\tt ammar.oulamara@loria.fr}.
   MACSI Project LORIA-INRIA Lorraine,
   Ecole des Mines de Nancy,
   Parc de Saurupt, 54042 Nancy, France.
   }
\and
Marie-Claude Portmann
   \thanks{ {\tt marie-claude.portmann@loria.fr}.
   MACSI Project LORIA-INRIA Lorraine,
   Ecole des Mines de Nancy,
   Parc de Saurupt, 54042 Nancy, France.
   }
}

\date{} 
\maketitle

\mbox{}
\vspace{-5ex}
\hrule height 1.5pt
\vspace{5ex}


\section{Introduction}
We consider the problem of minimizing the makespan in a flowshop involving maximal and minimal time lags, denoted by $Fm|\theta^{min},\theta^{max}|C_{max}$. Time lag constraints generalize the classical precedence constraints between operations: minimal (respectively maximal) time lags indeed specify that the time elapsed between two operations must be lower- (respectively upper-) bounded. We assume that such constraints are only defined between operations of the same job.

Time lags may be used to model various industrial situations. In several processes the delay between some critical operations must not be too long otherwise the products may be deteriorated. Such examples can be found in food industry with perishable products or in chemistry with unstable chemicals. Minimal time lags may correspond to transportation times between the machines or to communication delays between processors. Detailed case studies of shop problems involving time lags can be found in Deppner's thesis \cite{Freddy}.

From a computational complexity viewpoint, the existence of time lags make the problem NP-hard, even for special cases with two machines: Yu et al. \cite{yhl} showed that the two-machine flowshop problem with minimal time lags and unit processing times ($F2|p_{i,j}=1, \theta^{min}|C_{max}$ is strongly NP-hard. Besides, the two-machine flowshop with constant maximal time lags ($F2|\theta^{max}=\theta|C_{max}$) has been shown to be strongly NP-hard as well, even if we restrict the problem to permutation schedules \cite{fop}. 
  
\section{Solution approach}

In the remainder of the paper, we consider only permutation schedules, which are not dominant even for two-machine problems. Since the classical $Fm\pi| |C_{max}$ problem is strongly NP-hard for $m \geq 3$, the corresponding problem with time lags is strongly NP-hard as well. A branch-and-bound procedure has been proposed in \cite{fop} to solve this problem, when time lags are only defined between successive operations of the jobs. The algorithm makes use of a fundamental result which claims that the restricted problem of constructing an optimal schedule for a given job sequence is polynomial. 

We show that this result remains valid when time lags are defined between arbitrary couples of operations of the same jobs and when an arbitrary regular criterion is considered.

We propose also other extensions of the solution method:
\begin{itemize}
\item It is possible to take into account job release dates by adding a dummy initial machine with zero processing times and minimal time lags between this machine and the first real one equal to the release date.

\item By symmetry, tails can be introduced using a dummy final machine. The resulting problem is equivalent to minimizing the maximum lateness ($L_{max}$).
\end{itemize}  

Concernig the general flowshop problem, where non-permutation schedules are also taken into account, we define the restricted problem as constructing an optimal schedule for a given job sequence on first machine. We show that this problem is:
\begin{itemize}
\item polynomial for 2 machines with minimal time lags only

\item strongly NP-hard for 2 machines with maximal time lags 

\item strongly NP-hard for 3 machines with minimal time lags
\end{itemize}

Therefore the proposed approach can only be used for the 2-machine flowshop problem with minimal time lags only.


\begin{thebibliography}{99}  

\bibitem{Freddy}
{\sc F. Deppner} (2004).
Ordonnancement d'atelier avec contraintes temporelles entre opérations.
PhD thesis, Institut National Polytechnique de Lorraine, France (in French).

\bibitem{fop}
{\sc J. Fondrevelle, A. Oulamara and M.-C. Portmann}.
Permutation flowshop scheduling problems with maximal and minimal time lags.
{\em Computers and Operations Research}, (to appear).

\bibitem{yhl}
{\sc W. Yu, H. Hoogeveen and J.K. Lenstra} (2004).
Minimizing makespan in a two-machine flow shop with delays and unit-time operations is NP-hard.
{\em Journal of Scheduling}, 7, 333-348.

\end{thebibliography}
\end{document}